\documentclass{amsart}    
\usepackage{amssymb}    
\usepackage{latexsym}    
    
\newcommand{\ZZ}{{\mathbb Z}}    
\newcommand{\RR}{{\mathbb R}}    
    
\newcommand{\NN}{{\mathbb N}}

\newtheorem{theorem}{Theorem}       
\newtheorem{lemma}{Lemma}[section]       
\newtheorem{prop}[lemma]{Proposition}       
\newtheorem{coro}[lemma]{Corollary}       
\newtheorem{definition}{Definition}       
\newtheorem{remark}{Remark} 

\newcommand{\calw}{{\mathcal{W}}}

\newcommand{\OOmega}{(\Omega,T)}
\newcommand{\GL}{GL(2,\RR)}
\newcommand{\SL}{SL(2,\RR)}
\newcommand{\gammanull}{\{E\in \RR : \gamma(E)=0\}}

\newcommand{\zet}{(\mathcal{Z})} 
\newcommand{\pus}{(\mathcal{S})} 
\newcommand{\abs}{(\mathcal{A})}

\renewcommand\qedsymbol{$\Box$}

\newcounter{smalllist}

\begin{document}    
\title[Singular spectrum of Lebesgue measure zero]{Singular spectrum of Lebesgue measure zero  for one-dimensional quasicrystals}
    
\author[D.~Lenz]{ Daniel Lenz $\,^{ 1,2,}$* }\thanks{* This research was supported in part by THE ISRAEL SCIENCE
FOUNDATION (grant no. 447/99) and  by the Edmund Landau Center for Research in Mathematical Analysis and Related Areas, sponsored by the Minerva Foundation (Germany). 
 } 

\maketitle

\vspace{0.3cm}        
\noindent 
$^1$ Institute of Mathematics, The Hebrew University, Jerusalem 91904, Israel \\[0.1cm] 
$^2$  Fachbereich Mathematik, Johann Wolfgang Goethe-Universit\"at,        
60054 Frankfurt, Germany\\[0.2cm]
E-mail: \mbox{ dlenz@math.uni-frankfurt.de}\\[3mm]        

\begin{abstract} 
The spectrum of one-dimensional discrete Schr\"odinger operators associated to strictly ergodic dynamical systems is shown to    coincide with the set of zeros of the   Lyapunov exponent  if and only if the Lyapunov exponent exists uniformly. This is used to  obtain Cantor spectrum of zero Lebesgue measure for all aperiodic subshifts  with uniform  positive weights. This  covers,   in particular, all aperiodic subshifts arising from  primitive substitutions including  new examples as e.g. the Rudin-Shapiro substitution. 

Our investigation is not based on trace maps. Instead it relies  on an Oseledec type theorem due to  A. Furman and a  uniform ergodic  theorem due to the author.
 
\end{abstract}

\section{Introduction}\label{introduction}

This article is concerned with discrete random Schr\"odinger operators associated to minimal subshifts over a finite alphabet. This means we consider a family $(H_\omega)_{\omega\in \Omega}$ of operators acting on $\ell^2 (\ZZ)$ by

\begin{equation}\label{family}
(H_\omega u) (n) \equiv u(n+1)  + u(n-1) + \omega(n) u(n),
\end{equation}
where $\omega\in \Omega$ and $\OOmega$ is a subshift over the finite set  $A\subset \RR$.  Recall that $\OOmega$ is called a subshift (over $A$) if $\Omega$ is a closed subset of $A^{\ZZ}$, invariant under the shift operator $T:A^{\ZZ}\longrightarrow A^{\ZZ}$   given by $(T a) (n)\equiv a(n+1)$. Here, $A$ carries the discrete topology and $A^{\ZZ}$ is given the product topology. A subshift is called minimal if every orbit is dense. 
For minimal subshifts $\OOmega$,   there exists a set $\Sigma\subset \RR$ s.t. 
\begin{equation}\label{minimal}
\sigma( H_\omega)= \Sigma,\:\; \mbox{ for all } \:\; \omega \in \Omega,
\end{equation}
where we denote the spectrum of the operator $H$ by $\sigma(H)$ (cf. \cite{BIST,Len1}).

Operators of this type arise in the quantum mechanical treatment of quasicrystals (cf. \cite{Baa, Sen} for background on quasicrystals). Various examples of such operators  have been studied in recent years. The main examples can be divided in two classes. These classes are  given by  primitive substitution operators (cf. e.g.  \cite{Bel,BBG,BG,Dam3,Sut,Sut2}) and    Sturmian operators respectively more generally circle map operators (cf. e.g. \cite{BIST, Dam4, DL,DL1, HKS,JL1, Kam}). A recent survey can be found in  \cite{Dam}.

For these classes and  in fact for arbitrary operators  of type \eqref{family}  satisfying suitable ergodicity and aperiodicity conditions, one expects the following features:
\\ $\pus$  Purely singular spectrum; $\abs$   absence of eigenvalues; $\zet$  Cantor spectrum of Lebesgue measure zero.

Note that  $\pus$ combined with $\abs$ implies purely singular continuous spectrum and note also  that $\pus$ is a consequence of   $\zet$.
Let us  mention that $\pus$ is by now  completely established for all relevant subshifts due do recent results of Last/Simon \cite{LS} in combination with earlier results of Kotani \cite{Kot}. For discussion of $\abs$ and further details we refer the reader to the cited literature.

The aim of this article is to investigate  $\zet$  and to relate it to ergodic properties of the underlying subshifts. 

The property $\zet$ has been investigated for several models by a number of authors: For  the period-doubling substitution it was shown by  Bellissard/Bovier/Ghez  in \cite{BBG}. There, it is also shown for the Thue-Morse substitution (cf.  earlier work of Bellissard \cite{Bel} as well). The most general result for primitive substitutions so far has been obtained by Bovier/Ghez \cite{BG}. They can treat a rather large class of primitive substitutions (given by an algorithmically accessible condition) including the examples mentioned above as well as new examples as e.g. the binary non-Pisot substitution.  For Sturmian operators $\zet$ has been established   by  S\"ut\H{o}   in the golden mean case ($=$ Fibonacci substitution) \cite{Sut,Sut2} and, extending this work,   by Bellissard/Iochum/Scoppola/Testard in the general case \cite{BIST}. A different approach  has been developed in \cite{DL6} by Damanik and the author. In \cite{DL6}, this approach is used to recover $\zet$  in the Sturmian case. A suitably modified version of this  approach can also be used to establish the result  for a certain class of substitutions as shown by Damanik \cite{Dam2}. While this class it not as big as the class investigated in \cite{BG}, it  contains many prominent examples including those mentioned above. 

All these results rely on the technique of trace maps (cf.   \cite{AP,Cas} as well for study of trace maps). Most of the cited works tackle not only $\zet$ but also $\abs$ (cf. Section \ref{further} for further discussion).

A canonical starting point in the investigation of $\zet$ is the fundamental result of Kotani \cite{Kot} that the set $\gammanull$
has Lebesgue measure zero if $\OOmega$ is aperiodic. Here, $\gamma$ denotes the Lyapunov exponent (precise definition given below).  This reduces the problem $\zet$ to establishing the equality
\begin{equation}\label{fundamentalgleichung}
\Sigma=\gammanull.
\end{equation}
As do all other investigations of $\zet$ so far, our approach starts from \eqref{fundamentalgleichung}. Unlike the  earlier treatments mentioned above   our approach     does not rely on trace maps. Instead, we present a new  method, the cornerstones of which are the following:

\begin{itemize}

\item A strong type of Oseledec theorem by A. Furman \cite{Fur}.

\item A uniform ergodic theorem for a large class of subshifts by the author \cite{Len2}. 

\end{itemize}

This  ergodic setting allows us to show that equation \eqref{fundamentalgleichung} may not only be seen as a  consequence of hierarchical structures as the trace map methods suggest but can rather be considered to be  a consequence of certain  ergodic properties of the system. In fact, we are able to even characterize validity of \eqref{fundamentalgleichung} by an  ergodic type condition, namely uniform existence of the Lyapunov exponent.   

On the conceptual level, this  provides a  new prospective on equality \eqref{fundamentalgleichung}.   On the practical level it provides a soft argument for   $\zet$ for a large class of examples. This class contains all  linearly repetitive subshifts and therefore, in particular,  all primitive subsitutions. Thus,  it gives information on the Rudin-Shapiro substitution wich could not be derived earlier (cf. discussion in \cite{BG}).   
To summarize, our aims are 

\begin{itemize}
\item[($*$)] to characterize validity of \eqref{fundamentalgleichung}  by an  essentially ergodic property of the subshift viz by uniform existence of the Lyapunov exponent (Theorem \ref{main}),
\item[($**$)] to present a large class of subshifts satisfying this  property (Theorem \ref{existence}). 
\end{itemize}

Here, $(*)$ gives the new conceptual point of view of our treatment and $(**)$ gives a large class of examples.

\medskip

Let us also   point out that our  characterization of validity of  \eqref{fundamentalgleichung} is not confined to subshifts over finite alphabets but  applies to arbitrary strictly ergodic systems.

\medskip

The paper is organized as follows. In Section \ref{Notation} we present the subshifts we will be interested in, introduce some notation  and state our results. 
In Section \ref{key}, we recall   results of Furman \cite{Fur} and  of the author \cite{Len2} and adopt them to our setting. Section \ref{proof} is devoted to a proof of our results. Finally, in Section \ref{further} we provide some further comments. 

\medskip

{\it Note added.} After this work was completed, we learned about the very recent preprint ``Measure Zero Spectrum of a  Class of Schr\"odinger Operators'' by Liu/Tan/Wen/Wu (mp-arc 01-189). They present a  detailed and thorough analysis of trace maps for primitive substitutions. Based on this analysis, they establish $\zet$ for all primitive substitutions thereby extending the approach  developed in  \cite{BBG,BG,Cas,Sut}.

\section{Notation and  Results}\label{Notation}
In this section we  discuss basic material concerning subshifts and the associated operators and state our results.

\medskip

We begin with a short discussion of  subshifts.
We will consider the elements of $\OOmega$ as double sided infinite words and use notation and concepts from the theory of words.  To $\Omega$ we  associate thet set $\calw$ of  words associated to $\Omega$   consisting of all finite subwords of elements of $\Omega$. The length $|x|$ of a word $x\equiv x_1\ldots x_n$ with $x_j\in A$, $j=1,\ldots,n$, is defined by  $|x|\equiv n$. The number of occurences of $v\in \calw$ in $x\in\calw$ is denoted by $\sharp_v (x)$. 
A subshift is called uniquely ergodic if there exist only one normalized invariant measure on $\Omega$. It is said to be strictly ergodic (SE), if 
\begin{itemize}
\item[(SE)] the subshift is both uniquely ergodic and minimal.
\end{itemize}
A  minimal subshift is called aperiodic if there does not exist an $n\in \ZZ,$ $n  \neq 0$, and $\omega \in \Omega$ with $T^n\omega= \omega$.

Virtually all models for quasicrystals in one dimension considered so far are based on strictly ergodic systems. This applies in particular for the two classes of Sturmian and substitution subshifts mentioned above.    

Recently, a further  class of  strictly  ergodic subshifts has received attention, viz linearly repetitive ones. In fact, this class (and its  higher dimensional analog)  is put forward in a recent paper by Lagarias and Pleasants \cite{LP} as models for ``perfectly ordered quasicrystals''. In the one-dimensional case it has been investigated by  Durand  from a different point of view  including a characterization in terms of primitive $S$-adic systems \cite{Du2} (cf. \cite{DHS} as well). It  contains all subshifts arising from primitive substitutions as well as all those Sturmian systems whose rotation number has bounded  continued fraction as has e.g. the Fibonacci system \cite{LP,Len3}.

This class is particularly attractive as it is not given by a generating procedure but by a combinatorial condition. More precisely, a subshift is said to satisfy linear repetitivity (LR), if the following holds:
\begin{itemize}
\item[(LR)] There exists a $\kappa\in \RR$ with $\sharp_v (x)\geq 1$ whenever $|x|\geq \kappa |v|$ for $x,v\in \calw$.
\end{itemize}
It is true but not immediate from the definition  that (LR) implies strict ergodicity \cite{Du2,LP,Len2} (s. below as well).

We can now  introduce the class of subshifts we will be dealing with. They are those satisfying uniform positivity of weights (PW) given as follows:

\begin{itemize}
\item[(PW)] There exists a $C>0$ with $\liminf_{|x|\to \infty} \frac{\sharp_v (x)}{|x|} |v|\geq C$ for every $v\in \calw$. 
\end{itemize}
One might think of  (PW) as a strong type of minimality condition. Indeed, minimality can easily be seen to be equivalent to $ \liminf_{|x|\to \infty}  |x|^{-1} \sharp_v (x) |v|>0$ for every $v\in \calw$  \cite{Que}. 
To further  put this condition in prospective,  we mention that the following holds
$$ (LR) \Longrightarrow (PW) \Longrightarrow (SE).$$
The first implication is clear from the definitions (take $C\equiv \kappa^{-1}$). The second implication follows from results of the author \cite{Len2}. 
%Thus,  the class of subshifts satisfying (PW) is smaller than the class of st%rictly ergodic one%s. However, it does contain all ``perfectly ordered quasicr%ystals'' in the sense of \cite{LP}, %including all primitive substitutions and% Fibonacci like Sturmian systems.  
In our setting the class of  subshifts satisfying (PW) appears naturally as it is exactly the class of subshifts admitting  a strong form of uniform ergodic theorem \cite{Len2}. Such a theorem in turn  is needed to apply Furmans results (s. below for details). 

\medskip

After this discussion of background from dynamical systems let us now get back to spectral theoretic issues. An important tool in spectral theoretic considerations are transfer matrices and Lyapunov exponents. These quantities will be introduced next.  To have the appropriate setting for the discussion in Section \ref{key}, we will actually choose a rather general approach. 

Let $\GL$ be the group of invertible $2\times 2$-matrices over $\RR$ and let $\SL$ be the subgroup of $\GL$ consisting of matrices with determinant equal to one. The operator norm $\|\cdot\|$ on the set of $2\times 2$-matrices induces a topology on $\GL$ and $\SL$. For a continuous function $A: \Omega \longrightarrow \GL$, $\omega\in \Omega, $ and $n\in \ZZ$, we define the cocycle $A(n,\omega)$ by

\begin{equation*}
A(n,\omega) \equiv \left\{\begin{array}{r@{\quad:\quad}l}
 A(T^{n-1} \omega)\cdots A(\omega)  & n>0\\
 Id & n=0\\
A^{-1} (T^n \omega) \cdots A^{-1} (T^{-1}\omega)  & n < 0
\end{array}\right.
\end{equation*}

By Kingmans subadditive ergodic theorem (cf. e.g. \cite{KW}), there exists  $\Lambda(A)\in \RR$ with
$$ \Lambda (A)= \lim_{|n| \to \infty} \frac{1}{|n|} \log \| A(n,\omega)\|$$
for $\mu$ a. e. $\omega \in \Omega$ if $\OOmega$ is uniquely ergodic with invariant probability    measure $\mu$. 
Following \cite{Fur}, we introduce the following definition.

\begin{definition} Let $\OOmega$ be strictly ergodic.  The continuous function  $A: \OOmega \longrightarrow \GL$ is called uniform   if the limit  $\Lambda (A)= \lim_{|n| \to \infty} \frac{1}{|n|} \log \| A(n,\omega)\|$ exists  for all $\omega \in \Omega$ and the convergence is uniform on $\Omega$. 
\end{definition}

\begin{remark}{\rm  It is possible to show that uniform existence of the limit in the definition already implies uniform convergence. The author learned this from Furstenberg and Weiss \cite{FW}. They  actually have a  more general result. Namely,  they consider  a topological minimal dynamical system $\OOmega$ with compact metric space $\Omega$ and a continuous subadditive cocycle $(f_n)$ (i.e. $f_n$ are continuous real-valued  functions on $\Omega$ with $f_{n+m} (\omega)\leq f_n (\omega) + f_m ( T^n \omega)$ for all $n,m\in \NN$ and $\omega \in \Omega$).  Their result then gives that existence of $\phi (\omega)= \lim_{n\to \infty} n^{-1} f_n (\omega)$ for all $\omega \in \Omega$ implies constancy of $\phi$ as well as uniform convergence. 

The proof proceeds by exhibiting a  dense $G_\delta$-set  in $\Omega$  on which  both  $\phi (\omega) \leq \liminf_{n\to \infty} (\min\{n^{-1} f_n (\omega) : \omega \in \Omega\})$ and  $\phi (\omega) \geq \limsup_{n\to \infty} (\max\{n^{-1} f_n (\omega) : \omega \in \Omega\})$  hold.  }

\end{remark}

For spectral theoretic investigations  a special type of $\SL$-valued function is relevant. Namely, for $E\in \RR$, we define the continuous function $M^E : \Omega \longrightarrow \SL$ by 
\begin{equation}  \label{transfer}
M^E(\omega) \equiv\left( \begin{array}{cc} E-\omega(1)  & -1\\1 & 0 \end{array}   \right).   
\end{equation}   
It is easy to see that  a sequence $u$ is a solution of the difference equation
\begin{equation}\label{gleichung}
u(n+1) + u(n-1)  + (\omega(n) -E) u (n)= 0
\end{equation}
if and only if 
\begin{equation} \label{wichtig}
\left( \begin{array}{c}  u(n+1) \\  u(n) \end{array}   \right) = M^E (n,\omega)\left( \begin{array}{c}  u(1) \\  u(0) \end{array}   \right), \: n\in \ZZ.
\end{equation}

By the above considerations, $M^E$ gives rise to the average   $\gamma(E)\equiv \Lambda(M^E)$. This average  is called the Lyapunov exponent for the energy $E$. It measures the rate of exponential growth of solutions of \eqref{gleichung}. 

\medskip

\medskip

Now, we are in a position to state our main result.

\begin{theorem}\label{main} Let $\OOmega$ be a strictly ergodic subshift over a finite alphabet. Then the following are equivalent:\\
(i) The function $M^E$ is uniform for every $E\in \RR$.\\
(ii) $\Sigma= \gammanull$. \\
In this case the Lyapunov exponent $\gamma :\RR\longrightarrow [0,\infty)$ is continuous. If in this case  furthermore $\OOmega$ is aperiodic, then the spectrum $\Sigma$ is a Cantor set of Lebesgue measure zero. 
\end{theorem}

\begin{remark}\label{remar}{\rm   (a) This theorem links (ii) to ergodic properties of the subshift, thereby giving the new prospective on (ii) discussed in the introduction. 
\\
(b) As will be seen later on,   $M^E$ is always uniform for $E$ with $\gamma(E)=0$ and for $E\in \RR\setminus \Sigma$. From this point of view, the theorem essentially states that $M^E$ can not be uniform for $E\in \Sigma$ with $\gamma(E)>0$.  \\
(c) Continuity of the Lyapunov exponent can easily be infered from (ii) (though this does not seem to be in the literature). More precisely, continuity of $\gamma$ on $\gammanull$ is a consequence of subharmonicity. Continuity of $\gamma$ on $\RR\setminus \Sigma$ follows from the Thouless formula (see e. g.  \cite{CL} for discussion of subharmonicity and the Thouless formula). Below, we will show that continuity of $\gamma$  follows from (i) and this will be crucial in our proof of  (i) $\Longrightarrow $ (ii).\\
(d) The  theorem (except its last statement)  and the proof given below remain valid for arbitrary  strictly ergodic dynamical systems.  Here, a dynamical system $\OOmega$ is called strictly ergodic if $\Omega$ is a compact metric space and $T$ is a homeomorphism of $\Omega$ s.t. every orbit is dense (minimality) and there is only one normalized $T$-invariant measure on $\Omega$ (unique ergodicity).
Of course, in this case $\omega (n)$ has to be replaced by $f(T^n \omega)$ in 
\eqref{family}, \eqref{transfer} and  \eqref{gleichung}, where $f:\Omega\longrightarrow \RR$ is a continuous function.  }
\end{remark}

Having studied $(*)$ of the introduction in the above theorem, we will now state our result on $(**)$.

\begin{theorem}\label{existence}
If $\OOmega$ satisfy (PW),  the function $M^E$ is uniform for each $E\in \RR$. 
\end{theorem}

\begin{remark}{\rm (a) Uniform existence of the Lyapunov exponent is rather unusual.  This is, of course, clear from Theorem  \ref{main}. Alternatively, it is not hard to see directly that it already fails 
 for discrete almost periodic operators. Namely, it is known that the Almost-Mathieu-Operator with coupling bigger than 2 has uniform positive Lyapunov exponent (cf. \cite{Her1}). By a  deterministic version of the theorem of Oseledec (cf. Theorem 8.1 of \cite{LS} for example), this would force pure point spectrum for all these operators, if the Lyapunov exponent existed uniformly on the spectrum. However,  there are examples of such  Almost-Mathieu Operators without point spectrum    \cite{AS,JS}. \\
(b) Uniform existence of the Lyapunov exponent has been shown for primitive substitutions by Hof in \cite{Hof} based on earlier work of Geerse/Hof \cite{GH}.  For certain Sturmian subshifts it has been shown in \cite{DL2}. For arbitrary linearly repetitive systems it has been investigated in  the authors thesis  (cf. \cite{DL5, Len2, Len3} as well). The above theorem contains all these results.\\
(c) The theorem is a rather direct consequence  of the subadditive theorem of \cite{Len2}. }
\end{remark}

The two theorems allow one to infer  some interesting conclusions.

\medskip

As validity of (ii) is known for arbitrary Sturmian dynamical subshifts \cite{BIST, Sut} (cf. \cite{DL6} as well), we have the following  corollary of Theorem \ref{main}. 

\begin{coro}\label{umkehrsturm}
Let $(\Omega(\alpha),T)$ be a Sturmian dynamical system with rotation number $\alpha$. Then $M^E$ is uniform for every $E\in \RR$. 
\end{coro}

\begin{remark}{\rm So far uniformity of $M^E$ for Sturmian systems could only be established for rotation numbers with bounded continued fraction expansion \cite{DL2}. This set of rotation numbers has measure zero. Thus, the corollary considerably extends one of the two main results of \cite{DL2}. Moreover, the corollary is remarkable as it is known that a general type of uniform ergodic theorem (or equivalently (PW) \cite{Len2}) actually fails as soon as the continued fraction expansion of $\alpha$ is unbounded \cite{Len2,Len3}. }
\end{remark}

Combining Theorem \ref{main} and Theorem \ref{existence}, we find the following corollary.

\begin{coro}\label{zeromeasure} Let $\OOmega$ sastisfy (PW). Then $\Sigma=\{E\in \RR : \gamma(E)=0\}$. If $\OOmega$ is furthermore aperiodic, then 
$\Sigma$ is a Cantor set of Lebesgue measure zero.
\end{coro}

\begin{remark}{\rm For aperiodic $\OOmega$ satisfying  (PW), this gives an alternative proof of $\pus$.}
%(a) The first part of the theorem covers the case of periodic subshifts as well. Of c%ourse, this case is well known. \\
%(b)  If $\OOmega$ satisfies (PW) and is aperiodic, this gives an alternative proof fo%r $\pus$.}
\end{remark}
As discussed  above  primitive substitutions satisfy (PW) (and even (LR)).  As validity of $\zet$ for primitive Substitutions has been a particular focus of earlier investigations, we explicitely state the following consequence of the foregoing corollary.

\begin{coro}
Let $\OOmega$ be aperiodic and  associated to a primitive substitution, then $\Sigma$ is a Cantor set of  Lebesgue measure zero.
\end{coro}

\begin{remark}{\rm  As discussed in the introduction, zero measure spectrum for substitutions has been considered  by several  authors \cite{BBG,BG,Dam2,Sut2}). The most general result has been the result of Bovier/Ghez \cite{BG}. They require existence of a square as well as a trace map satisfying a  semifiniteness condition \cite{BG}.  This is not satisfied by every substitution.  In particular, they could not treat the Rudin-Shapiro substitution (cf.  Section \ref{further} for further discussion). 
 } 
\end{remark}

\section{Key results} \label{key}
In this section, we present  results of Furman \cite{Fur} and  of the author \cite{Len2} and adopt them to our setting.  In the sequel  $\OOmega$ will always be a subshift over a finite alphabet as discussed in the introduction. However, let us emphasize that the results and proofs given below (with the only exception of Lemma \ref{set}) are valid  for arbitrary   dynamical systems over metric compact spaces (cf. Remark \ref{remar} (d)).

\medskip

We start with  some simple facts concerning uniquely ergodic systems. 
Define  for a continuous $b : \Omega \longrightarrow \RR$  and $n\in \ZZ$ the averaged function  $A_n(b) : \Omega\longrightarrow \RR$  by
\begin{equation}
A_n (b) (\omega) \equiv \left\{\begin{array}{r@{\quad:\quad}l}
 n^{-1}  \sum_{k=0}^{n-1} b( T^k \omega) & n>0\\
  0 & n=0\\
|n|^{-1} \sum_{k=1}^{|n|} b(T^{-k}\omega) & n < 0
\end{array}\right.
\end{equation}

The following proposition is well known see e.g. \cite{Wal}.   
\begin{prop}\label{ue} Let $\OOmega$ be  uniquely ergodic  with invariant probability  measure $\mu$. Let $b$ be a continuous function on $\Omega$.  Then the averaged functions $A_n (b)$ converge uniformly towards the constant function with value $\mu(b)$ for $|n|$ tending to infinity. 
\end{prop}

The following result by A. Furman  is crucial to our approach. It describes the structure of uniform functions with positive average. It can be seen as a continuous version of an Oseledec type theorem.

\begin{lemma}\label{structure} Let $\OOmega$ be strictly ergodic with invariant probability measure $\mu$. Let $B: \Omega \longrightarrow \SL$ be uniform with $\Lambda(B)>0$. Then, there exists a continuous function $G:\Omega\longrightarrow \GL$ and a continuous function $b:\Omega \longrightarrow \RR$ with $\mu(b) = \Lambda(B)$ such that for every $n\in \ZZ $ the following holds:
$$
B(n,\omega) = G(T^{n} \omega)^{-1}  
\left( \begin{array}{cc} \exp( - n A_n (b)(\omega))  & 0 \\ 0  & \exp (n A_n (b)(\omega)) \end{array}   \right)
G(\omega).
$$
\end{lemma}
{\it Proof.}  It suffices to show the result for $n=1$. The other cases then follow by multiplication and inversion.  Now, the result is essentially  given by   Theorem 4  in \cite{Fur}.  
While this  is quite clear, it is not completely  explicit from  the actual statement of the theorem. Thus, for the convenience of the reader and as we will use a similar reasoning later on, we include a discussion.

Theorem 4  of \cite{Fur}   states  that uniformity of $B$ implies  that (in the notation of \cite{Fur}) either   $\Lambda(B)=0$ or  $B$ is continuously diagonalizable. As we have $\Lambda(B)>0$, we infer that $B$ is continuously diagonalizable. This means that  there exist continuous functions  $C:\Omega\longrightarrow \GL $ and $a,d :\Omega \longrightarrow \RR$ with

\begin{equation}\label{eins}
B(1,\omega)= C(T \omega)^{-1} \left( \begin{array}{cc} \exp(a(\omega))  & 0 \\ 0  & \exp( d(\omega))\end{array}   \right) C(\omega).
\end{equation}
Setting, $h(\omega)\equiv \frac{1}{2} \log |\det C(\omega)|$ and $G(\omega)\equiv |\det C(\omega)|^{\frac{-1}{2}} C(\omega)$, we can rewrite $B(1,\omega)$  as 
$$
G(T \omega)^{-1}   \left( \begin{array}{cc}\exp ( a(\omega) -h(T\omega) +h(\omega)) & 0 \\ 0  &   \exp( d(\omega)  -h(T \omega) +h(\omega)   )  \end{array}   \right) G(\omega).
$$
Using $1=|\det G|= |\det B|$, we find 
$$ a(\omega) -h(T \omega) +h(\omega)= -( d(\omega)  -h(T \omega) +h(\omega)  )$$
Thus, defining $b:\Omega \longrightarrow \RR$ by $b(\omega)\equiv d(\omega)  -h(T^{-1}\omega) +h(\omega) $, we arrive at the desired equation
\begin{equation}\label{zwei}
B(1,\omega) = G(T  \omega)^{-1}  
\left( \begin{array}{cc} \exp( -  b(\omega))  & 0 \\ 0  & \exp( b(\omega)) \end{array}   \right)
G(\omega).
\end{equation}

It remains to show the statement about $\mu(b)$. By compactness of $\Omega$ and continuity of $G$ we have $\Lambda(B)= \Lambda( G (T\cdot) B G^{-1})$. Using this, we infer from 
\eqref{zwei} and the previous Proposition 
  $$\Lambda(B) = \Lambda( \left( \begin{array}{cc} \exp( -  b(\omega))  & 0 \\ 0  & \exp( b(\omega)) \end{array}   \right)    ) =  |\mu(b)|.$$
 If $\mu(b) = |\mu(b)|$ the proof is finished. If $\mu(b) = - |\mu(b)|$, the claim follows after a suitable conjugation. \hfill \qedsymbol

\begin{lemma}\label{continuitylemma} 
Let $\OOmega$ be strictly ergodic. Let $A: \Omega \longrightarrow \SL$ be uniform. Let $(A_n)$ be a sequence of  continuous $\SL$-valued functions converging to $A$ in the sense that $d(A_n, A) \equiv \sup_{\omega \in \Omega} \{ \|A_n (\omega) -A(\omega)\|\}\longrightarrow 0$, $n\longrightarrow \infty$. Then, $\Lambda(A_n) \longrightarrow \Lambda(A)$, $n\longrightarrow \infty$. 
\end{lemma}
{\it Proof.} Again this is essentially a result of \cite{Fur}.  More precisely,   Theorem 5 of \cite{Fur} shows that  $\Lambda(A_n)$ converges to $\Lambda(A)$ whenever the following holds:  $A$ is a uniform  $\GL$-valued function  and $d(A_n,A)\longrightarrow 0$  and $d(A_n^{-1},A^{-1})\longrightarrow 0$, $n\longrightarrow \infty$.  Now, for functions $A_n,A $ with values in $\SL$, it is easy to see that  $d(A_n^{-1},A^{-1})\longrightarrow 0$, $n\longrightarrow \infty$ if  $d(A_n,A)\longrightarrow 0, n\longrightarrow \infty$. The proof of the lemma is finished. \hfill \qedsymbol

\medskip

\begin{lemma}\label{null} Let $\OOmega$ be strictly ergodic. Let $A:\Omega\longrightarrow \GL$ be continuous.  Then, the inequality $\limsup_{n\to \infty} n^{-1} \log \| A(n,\omega)\| \leq \Lambda(A)$ holds uniformly on $ \Omega$. 
\end{lemma}
{\it Proof.} This follows from Theorem 1 of \cite{Fur}. \hfill \qedsymbol

\medskip

Finally, we need the following lemma providing a large supply of uniform functions if $\OOmega$ satisfies (PW). 

\begin{lemma}\label{set} Let $\OOmega$ satisfy (PW). Let $F: \calw\longrightarrow \RR$ satisfy  $F(xy) \leq F(x) + F(y)$ (i.e. $F$ is subadditive). Then, the limit $\lim_{|x|\to \infty} \frac{F(x)}{|x|}$ exists. 
\end{lemma}
{\it Proof.} This is just one half of Theorem 2 of \cite{Len2}. \hfill \qedsymbol

\section{Proofs of the main results}\label{proof}

In this section, we use the results of the foregoing section to prove the theorems stated in Section \ref{Notation}.  Again let us point out that the lemmas given below  and their proofs  are valid for arbitrary topological dynamical systems over metric compact spaces (cf. Remark \ref{remar} (d)).

\medskip

We start with some lemmas needed for the proof of Theorem \ref{main}.  We will denote the Euclidean norm of an element $v\in \RR^2$ by $\|v\|$ i.e. $\|v\|^2\equiv v_1^2 + v_2 ^2$ for $v$ with components $v_1$ and $v_2$. 

\begin{lemma} \label{richtungeins}Let $\OOmega$ be strictly ergodic. If $M^E$ is uniform for every $E\in \RR$ then $\Sigma=\gammanull$ and  the Lyapunov exponent $\gamma: \RR \longrightarrow [0,\infty)$ is continuous. 
\end{lemma}

{\it Proof.} We start by showing continuity of the Lyapunov exponent.  Consider a sequence $(E_n)$ in $\RR$ converging to $E\in \RR$. As the function $M^E$ is uniform by assumption,   by Lemma \ref{continuitylemma}, it suffices to show that $d(M^{E_n},M^E)\rightarrow 0, n\rightarrow \infty$. This is clear  from the definition of $M^E$ in \eqref{transfer}.

\medskip

Set $\Gamma\equiv \gammanull$.   The inclusion $\Gamma \subset \Sigma$ follows from general principles (cf. e.g. \cite{CL}). Thus, it suffices to show the opposite inclusion $\Sigma \subset \Gamma$. By \eqref{minimal}, it suffices to show $\sigma(H_\omega) \subset \Gamma$ for a fixed $\omega\in \Omega$.  

Assume the contrary. Then there exists spectrum of $H_\omega$ in the complement $\Gamma^c\equiv \RR \setminus \Gamma$ of $\Gamma$ in $\RR$. As $\gamma$ is continuous,  the set $\Gamma^c$ is open. Thus,   spectrum of $H_\omega$  can only  exist in $\Gamma^c$, if spectral  measures of $H_\omega$ give actually weight to   $\Gamma^c$. By standard  results on generalized eigenfunction  expansion \cite{Ber}, there exists then  an $E\in \Gamma^c$  admitting a polynomially bounded solution $u\neq 0$ of \eqref{gleichung}. This gives a contradiction in the following way:

By \eqref{wichtig} and Lemma \ref{structure} applied to $M^E$ (use $E\in \Gamma^c$ to obtain $\Lambda(M^E)\equiv \gamma(E)>0$),  there exist continuous functions $G:\Omega \longrightarrow \GL$ and $b:\Omega \longrightarrow \RR$ with $\mu(b) = \Lambda(M^E)=\gamma(E)>0$ with 
\begin{equation}\label{haha}
%(u(n+1), u(n))^t
\left( \! \begin{array}{c}  u(n+1) \\  u(n) \end{array}  \! \right)
= G(T^n \omega)^{-1} 
\left(\! \begin{array}{cc} \exp( - n  A_n (b)(\omega) )  & 0 \\ 0  & \exp ( n  A_n (b)(\omega)) \end{array}   \!\right)
G(\omega)
\left(\! \begin{array}{c}  u(1) \\  u(0) \end{array}  \! \right)
% (u(1), u(0))^t
\end{equation}
for every $n\in \ZZ$. 
As $G:\Omega \longrightarrow  \GL$ is continuous  on the compact space $\Omega$, there exist constants $\rho_1,\rho_2$, $0<\rho_1,\rho_2 <\infty$,  with
\begin{equation}\label{bounds}
\|G(\omega)\| \leq \rho_1, \:\;\: \|G(\omega)^{-1} v\| \geq \rho_2 \|v\|
\end{equation}
for every $\omega \in \Omega$ and every $v\in \RR^2$. Moreover,  
as $b$ is continuous and $\OOmega$ is strictly ergodic, we infer from Proposition \ref{ue}
\begin{equation} \label{mittel} \lim_{|n|\to \infty} A_n (b)(\omega)  = \mu(b)=\Lambda(M^E) = \gamma(E) >0.
\end{equation}

Set $\left( \begin{array}{c}  x \\  y \end{array}   \right) \equiv G(\omega )\left( \begin{array}{c} u(1) \\  u(0) \end{array}   \right)$.
As $u$ is polynomially bounded, we infer $y=0$ from 
\eqref{haha}, \eqref{bounds} and \eqref{mittel} by considering the right half-axis i.e.   large values of $n$. Similarly, considering the left half-axis i.e. small values of $n$, we infer $x=0$. As $G(\omega)$ is invertible this means $u(0)=u(1)=0$ and this gives the contradiction  $u\equiv 0$.   \hfill \qedsymbol

\begin{lemma}\label{nullstellenmenge} If $\OOmega$ is strictly ergodic,  $M^E$ is uniform for each $E\in \RR$  with $\gamma(E)=0$.
\end{lemma}
{\it Proof.} By $\det M^E (\omega)= 1$, we have $1\leq \|M^E(n, \omega)\|$ and therefore $0\leq \liminf_{n\to \infty}  n^{-1} \log \|M^E(n,\omega)\| \leq \limsup_{n\to \infty}  n^{-1} \log \|M^E(n,\omega)\|$. Now, the statement follows from Lemma \ref{null}. \hfill \qedsymbol

\medskip

Moroeover, we have the following lemma. The lemma is certainly well known. However, we could not find a proof in the literature. Thus, for the convenience of the reader, we include a proof. 

\begin{lemma}\label{resolvente} If $\OOmega$ is strictly ergodic,  $M^E$ is uniform  with $\gamma(E)>0$ for each $E\in \RR\setminus \Sigma$.
\end{lemma}
{\it Proof.} Let $E\in \RR\setminus \Sigma$ be given.  The proof will be split in four steps. Recall that $\Sigma$ is the spectrum of $H_\omega$ for every $\omega \in \Omega$ by \eqref{minimal} and thus $E$ belongs to the resolvent of $H_\omega$ for all $\omega\in \Omega$. 

\medskip

{\it Step 1}. For every $\omega\in \Omega$, there exist unique (up to a sign) normalized  $U(\omega), V(\omega)\in \RR^2$ such that  $\|M^E (n,\omega) U(\omega)\|$ is exponentially decaying for $n\longrightarrow \infty$ and $\|M^E (n,\omega) V(\omega)\|$ is exponentially decaying for $n\longrightarrow -\infty$. The vectors $U(\omega),V(\omega)$ are linearly independent. For fixed $\omega\in \Omega$ they can be choosen to be continuous in a neighborhood of $\omega$. 

\medskip

{\it Step 2}. Define the matrix $C(\omega)$ by $C(\omega)\equiv (U(\omega),V(\omega))$. Then $C(\omega)$ is invertible and there exist functions $a , b : \Omega\longrightarrow  \RR\setminus \{0\}$ such that
\begin{equation} \label{diagonal} C(T\omega)^{-1} M^E (\omega)  C(\omega) = \left(\! \begin{array}{cc} a (\omega)  & 0 \\ 0  &  b(\omega) \end{array}   \!\right).
\end{equation}

\medskip

{\it Step 3}. The functions $|a|, |b|, \|C\|, \|C^{-1}\| : \Omega\longrightarrow \RR$ are continuous. 

\medskip

{\it Step 4}. $M^E$ is uniform with $\gamma(E)>0$.

\medskip

Ad Step 1. This is standard up to the continuity statement. Here is a sketch of the construction:  Fix $\omega \in \Omega$. Set  $u_0 (n)\equiv (H_\omega -E)^{-1} \delta_0 (n)$ and $u_{-1} (n)\equiv (H_\omega -E)^{-1} \delta_{-1} (n)$, where  $\delta_k$ $, k\in \ZZ$,  is the sequence  in $\ell^2 (\ZZ)$  vanishing everywhere except in $k$, where it takes the value $1$.   By Combes/Thomas arguments, see e.g. \cite{CL}, the vectors $(u_0 (0), u_0 (1))^t$ and $ (u_{-1} (0), u_{-1} (1))^t$  give rise to solutions of \eqref{gleichung} wich decay exponentially for $n\to \infty$.  Here, $(x,y)^t$ denotes the transpose of $(x,y)$. It is easy to see that not both of these solutions can vanish identically. Thus, after normalizing, we find a vector $U(\omega)$ with the desired properties. By continuity of $\omega \mapsto (H_\omega  - E)^{-1} \delta_k$, $k\in \ZZ$, we can choose $U(\omega)$ continuous in a neighborhood of a fixed $\omega$. The construction for $V(\omega)$ is similar focusing on behaviour of solutions to the left i.e. for $n\longrightarrow -\infty$. 

Uniqueness follows by standard arguments from constancy of the Wronskian. Linear independence is clear as $E$ is not an eigenvalue of $H_\omega$. 

\medskip

Ad Step 2.  The matrix $C$ is invertible by linear independence of $U$ and $V$. The uniqueness statements of Step 1, show that there exist functions $a,b:\Omega\longrightarrow \RR$ with $M^E (\omega ) U(\omega) = a(\omega) U(T\omega)$ and $M^E (\omega) V(\omega) = b(\omega) V(T\omega)$. This easily yields \eqref{diagonal}. As the left hand side of this equation is invertible, the right hand side is invertible as well. This shows that $a$ and $b$ do not vanish anywhere. 

\medskip

Ad Step 3. Direct calculations show that the functions in question do not change if $U(\omega)$ or $V(\omega)$ or both are replaced by $-U(\omega) $ resp. $-V(\omega)$. By Step 1,  such a replacement can be used to provide a  version of $V$ and $U$ continuous arround an arbitrary $\omega\in \Omega$. This gives the desired  continuity.

\medskip

Ad Step 4. As $\|C\|$ and $\|C^{-1}\|$ are continuous by Step 3 and $\Omega$ is compact, there exist constants $\kappa_1, \kappa_2$ with $0<\kappa_1, \kappa_2<\infty$ with $\kappa_1 \leq \|C(\omega)\|, \|C^{-1} (T \omega)\|\leq \kappa_2$ for every $\omega \in \Omega$. Thus, uniformity of $M^E$ will follow from uniformity of $\omega \mapsto C^{-1} (T\omega) M^E (\omega) C (\omega)$, which in turn will follow by Step 2 from uniformity of
$$ \omega \mapsto D(\omega)\equiv \left(\! \begin{array}{cc} |a| (\omega)  & 0 \\ 0  &  |b|(\omega) \end{array}   \!\right).$$
As $|a|$ and $|b|$ are continuous by Step 3 and do not vanish by Step 2, the functions $\ln |a|, \ln |b| : \Omega \longrightarrow \RR$ are continuous. The desired uniformity of $D$ follows now by Proposition \ref{ue} (see proof of Lemma \ref{structure}
 for a similar reasoning). Positivity of $\gamma(E) $ is immediate from Step 1. \hfill \qedsymbol

\medskip

A  simple but crucial step in the proof of Theorem \ref{existence} is to relate the transfer matrices to subadditive functions. This
 will allow us to use Lemma \ref{set} to  show that the uniformity  assumption of Lemma \ref{structure} and Lemma \ref{continuitylemma} holds  for subshifts  satisfying (PW). We proceed as follows. 
Let $\OOmega$ be a strictly ergodic subshift and let $E\in \RR$ be given. To the matrix valued function $M^E$ we associate the function $F^E: \calw \longrightarrow \RR$ by setting

$$ F^E (x)\equiv \log \| M^E(|x|, \omega)\|,$$
where $\omega\in \Omega$ is arbitrary with $\omega(1)\ldots \omega(|x|)=x$. It is not hard to see that this is well defined. Moreover, by submultiplicativity of the norm $\| \cdot\|$, we infer that $F^E$  satisfies $F^E(xy)\leq F^E (x) + F^E (y)$.

\begin{prop}\label{useful} $M^E$ is uniform if and only if the  limit $\lim_{|x|\to \infty} \frac{F^E(x)}{|x|}$ exists.
\end{prop}
{\it Proof.} This is straightforward. \hfill \qedsymbol

\medskip

Now, we can prove the results stated in Section \ref{Notation}.

\medskip

{\it Proof of Theorem \ref{main}.} The implication (i)$\Longrightarrow$(ii)  is an immediate consequence of Lemma \ref{richtungeins}. This lemma also shows continuity of the Lyapunov exponent. The implication (ii)$\Longrightarrow$(i)     follows from Lemma \ref{nullstellenmenge} and Lemma \ref{resolvente}. 

It remains to show the statement on aperiodic subshifts.  As $\Sigma$ is closed and has no discrete points by general principles on random operators, the Cantor property will follow if $\Sigma$ has measure zero. But this follows from the first statement of the theorem, as  the set $\gammanull$ has measure zero by the results of Kotani theory discussed in the introduction.

\hfill \qedsymbol

\medskip

{\it Proof of Theorem \ref{existence}.} This is immediate from Lemma \ref{set} and Proposition \ref{useful}. \hfill \qedsymbol

\section{Further discussion}\label{further}

In this section we will present some discussion and comments on the results proven in the previous sections. 

As shown in the introduction and the proof of Theorem \ref{main}, the problem $\zet$  can essentially be reduced to establishing the inclusion $\Sigma \subset \gammanull.$
 This has been investigated for various models by various authors \cite{BBG, BIST,BG,Dam2,DL6,Sut2}. All these  proofs  rely on the same tool viz    trace maps.  Trace maps are very   powerful as they  capture the underlying  hierarchical structures.  
Besides beeing applicable in the investigation of $\zet$,  trace maps are extremely useful because 
\begin{itemize}
\item trace map bounds are an important tool to   prove absence of eigenvalues.
\end{itemize} 
As already mentioned,  most of the cited literature actually studies both $\abs$ and $\zet$. In fact,    $\zet$ can even be shown to follow from a strong version of $\abs$ \cite{DL6} (cf. \cite{Dam2} as well). 
While this makes the trace map approach to $\zet$  very attractive, it has  two drawbacks: 
\begin{itemize}
\item The analysis of  the actual trace maps  may be quite  hard or even impossible. 

\item  The trace map formalism only applies to substitution-like subshifts. 
\end{itemize}

Thus, trace map methods can not be expected to   establish zero-measure spectrum  in  a generality comparable to the validity of the underlying Kotani result. 

\medskip

Let us now  compare this with the method presented above. Essentially, our method has a  complementary profile:  It does not seem to  give  information concerning absence of eigenvalues. But on the other hand it only requires a weak ergodic type condition. 

This condition is in particular met by subshifts satisfying (PW). This class of subshifts  is rather large.   It contains all  linearly repetitive subshifts (``perfectly ordered quasicrystals'' in the sense of \cite{LP}) and a fortiori all primitive substitutions. In particular, it gives information on the Rudin-Shapiro substitution which so far had been  unattainable.
Moreover, quite likely, the condition (PW) will be satisfied for certain   circle maps, where $\zet$ could not be proven by other means   (except of course Sturmian ones s. below). 

Thus, (PW) provides  a natural framework for validity of $\zet$ providing a large class of examples. 

All the same,  it seems worthwhile pointing out that (PW)  does  not contain the class of Sturmian systems whose rotation number has unbounded continued fraction expansion. This is in fact the only class known to satisfy $\zet$ (and much more \cite{BIST,Dam4, DL,DL1,DL2, JL1,JL2,Sut})  not covered by   (PW). For this class, one can use the implication (ii) $\Longrightarrow$ (i) of Theorem \ref{main}, to conclude uniform existence of the Lyapunov exponent as done in Corollary \ref{umkehrsturm}. Still it seems desirable to give a direct proof of uniform existence of the Lyapunov exponent for these systems. 

\medskip

Finally, let us briefly discuss the case of operators associated to arbitrary strictly ergodic dynamical systems (cf. Remark \ref{remar} (d)). As discussed above, Theorem \ref{main} remains valid in this case. But it might be of rather limited use. Thus, the following result (which is essentially a corollary of the proof of Theorem \ref{main}) might be a  more appropriate formulation in this case. 

\begin{theorem} Let $\OOmega$ be an arbitrary strictly ergodic dynamical system. Then, $\RR\setminus \Sigma$ is open and $M^E$ is uniform with $\gamma(E)>0$ for every $E\in \RR\setminus \Sigma$. Conversely, if $I\subset \RR$ is open such that $M^E$ is uniform with $\gamma(E)>$ for every $E\in I$, then $I$ is contained in $\RR\setminus \Sigma$. 
\end{theorem}
{\it Proof. } Openess of $\RR\setminus \Sigma$ is clear. Now, the first statement follows from  Lemma \ref{resolvente}. The second statement follows from the proof of Lemma \ref{richtungeins} after replacing $\Gamma^c$ by $I$. \hfill \qedsymbol

\medskip

The theorem characterizes the  complement of the spectrum in $\RR$     as the largest open set in $\RR$ on which uniformity and positivity of the Lyapunov exponent hold.  \\[3mm]

{\footnotesize {\it Acknowledgements.} This  work  was done while the author was visiting The Hebrew University, Jerusalem. 
The author would like to thank  Y. Last for hospitality as well as for many stimulating conversations on a wide range of topics including those considered above. Enlightening discussions with B. Weiss are also gratefully acknowledged.  Special thanks are due to H. Furstenberg for most valuable discussions and for bringing  the work of A. Furman   \cite{Fur} to the authors  attention. The author would also like to thank D. Damanik for an earlier collaboration on the topic of zero measure spectrum \cite{DL6}.
% Financial support from both The Edmund Landau Center for Research in Analysis and The Israel Science Founda%tion is gratefully acknowledged. 
}


\begin{thebibliography}{10}  

\bibitem{AP} J.-P. Allouche, J. Peyi\`{e}re, Sur une formule de r\'{e}currence sur les traces de produits de matrices associ\'{e}s \`{a} certaines substitutions, \textit{C.R. Acad. Sci. Paris} {\bf 302} (1986), 1135--1136

\bibitem{AS}  J. Avron,  B. Simon, Almost periodic  Schr\"odinger Operators,  II.
The integrated density of states, \textit{  Duke Math. J.}, {\bf 50} (1983), 369--391
                  
\bibitem{Baa} M. Baake, A guide to mathematical quasicrystals, in: \textit{Quasicrystals}, Eds. J.-B. Suck, M. Schreiber, P. H\"aussler, Springer, Berlin (1999)

\bibitem{Bel} J. Bellissard, Spectral properties of Schr\"odinger operators with a Thue-Morse potential, in: \textit{ Number theory and physics},  Eds. J.-M. Luck, P.  Moussa, M.  Waldschmidt, Proceedings in Physics, {\bf 47}, Berlin, Springer (1989), 140--150                  


\bibitem{BBG} J.\ Bellissard, A.\ Bovier,  J.-M.\ Ghez, Spectral properties of a tight binding Hamiltonian with period doubling potential,
\textit{Commun.\ Math.\ Phys.} {\bf 135} (1991), 379--399

\bibitem{BIST} J.\ Bellissard, B.\ Iochum, E.\ Scoppola, and D.\ Testard, Spectral properties of one-dimensional quasi-crystals,
\textit{Commun.\ Math.\ Phys.} {\bf 125} (1989), 527--543

\bibitem{BG} A. Bovier, J.-M. Ghez, Spectral Properties of One-Dimensional Schr\"odinger Operators with Potentials Generated by Substitutions, \textit{Commun. Math. Phys.} {\bf 158} (1993), 45--66; Erratum: Commun. Math. Phys. {\bf 166}, (1994), 431--432

\bibitem{Ber} J. M.   Berezanskii, Expansions in eigenfunctions of self-adjoint operators, Transl. Math. Monographs {\bf 17}, Amer. Math. Soc. Providence,  R.I. (1968)

\bibitem{Cas} M. Casdagli, Symbolic dynamics for the renormalization map of a quasiperiodic Schr\"odinger equation, \textit{Commun. Math. Phys.}  {\bf 107} (1986), 295--318


\bibitem{CL} R. Carmona, J. Lacroix, \textit{Spectral theory of Random Schr\"odinger Operators}, Birkh\"auser, Boston (1990)


\bibitem{Dam3} D.\ Damanik, Singular continuous spectrum for a class of substitution Hamiltonians, \textit{Lett.\ Math.\ Phys.} {\bf 46} (1998),
303--311


\bibitem{Dam4} D. Damanik, $\alpha$-continuity properties of one-dimensional quasicrystals, \textit{Commun.\ Math.\ Phys.} {\bf 192} (1998),
169--182

\bibitem{Dam2} D.\ Damanik, Substitution Hamiltonians with bounded trace map orbits, \textit{J.\ Math.\ Anal.\ Appl.} {\bf 249} (2000), 393--411


\bibitem{Dam} D. Damanik, Gordon-type arguments in the spectral theory of one-dimensional quasicrystals, in: {\it Directions in Mathematical
Quasicrystals}, Eds.~M.~Baake, R.~V.~Moody, CRM Monograph Series {\bf 13}, AMS, Providence, RI (2000), 277--305



\bibitem{DL} D.\ Damanik, R.\ Killip, and D.\ Lenz, Uniform spectral properties of one-dimensional quasicrystals, III. $\alpha$-continuity,
\textit{Commun.\ Math.\ Phys.} {\bf 212} (2000), 191--204 

\bibitem{DL1} D.\ Damanik,  D.\ Lenz, Uniform spectral properties of one-dimensional quasicrystals, I. Absence of eigenvalues,
\textit{Commun.\ Math.\ Phys.} {\bf 207} (1999), 687--696

\bibitem{DL2} D.\ Damanik, D.\ Lenz, Uniform spectral properties of one-dimensional quasicrystals, II. The Lyapunov exponent, \textit{Lett. Math. Phys.}  {\bf 50} (1999), 245--257

\bibitem{DL5} D.  Damanik, D. Lenz,   Linear repetitivity I., Subadditive
  ergodic theorems,  to appear in: \textit{Discr. Comput. Geom.}


\bibitem{DL6} D. Damanik, D. Lenz, Half-line eigenfunctions estimates and singular continuous spectrum of zero Lebesgue measure, preprint

%\bibitem{DP} F.\ Delyon and D.\ Petritis, Absence of localization in a class of Schr\%"odinger operators with quasiperiodic potential,
%\textit{Commun.\ Math.\ Phys.} {\bf 103} (1986), 441--444


\bibitem{Du2} F. Durand, Linearly recurrent subshifts have a finite
  number of non-periodic subshift factors, {\it Ergod. Th. \&
    Dynam. Sys.} {\bf 20} (2000), 1061--1078

\bibitem{DHS} F. Durand, F.  Host,  C.   Skau, Substitution dynamical
  systems, Bratteli diagrams and dimension groups, {\it Ergod. Th.\&
    Dynam. Sys.} {\bf 19} (1999),  953--993 

\bibitem{Fur} A. Furman, On the multiplicative ergodic theorem for uniquely ergodic ergodic systems, \textit{Ann. Inst. Henri Poincar\'{e} Probab. Statist.} {\bf 33} (1997), 797--815

\bibitem{FW} H. Furstenberg, B. Weiss, private communication

\bibitem{GH} C. Geerse, A.  Hof, Lattice gas models on self-similar
  aperiodic tilings, {\it Rev. Math. Phys.} {\bf 3} (1991), 163--221

\bibitem{Her1}  M.-R. Herman, Une m\'{e}thode pour minorer les exposants de Lyapunov et quelques exemples montrant the caract\`{e}re local d'un  th\'{e}or\`{e}me d'Arnold et de Moser sur le tore de dimension $2$, \textit{Comment. Math. Helv} {\bf 58} (1983), 4453--502

\bibitem{Hof} A.  Hof,  Some Remarks on Aperiodic Schr\"odinger
  Operators, {\it J. Stat. Phys.} {\bf 72} (1993), 1353--1374 

\bibitem{HKS} A.\ Hof, O.\ Knill, B.\ Simon, Singular continuous spectrum for palindromic Schr\"odinger operators, \textit{Commun.\ Math.\
Phys.} {\bf 174} (1995), 149--159


\bibitem{JL1} S. Jitomirskaya, Y. Last, Power law subordinacy and singular spectra. I. Half-line operators, \textit{Acta Math.} {\bf 183} (1999), 171--189

\bibitem{JL2} S. Jitomirskaya, Y. Last, Power law subordinacy and singular spectra. II. Line Operators, \textit{Commun. Math. Phys.} {\bf 211} (2000), 643--658

\bibitem{JS} S. Jitomirskaya, B. Simon,  Operators with singular continuous spectrum. III. Almost perodic Schr\"odinger operators, \textit{Commun. Math. Phys.}, {\bf 165} (1994), 201--205

\bibitem{Kam} M.\ Kaminaga, Absence of point spectrum for a class of discrete Schr\"odinger operators with quasiperiodic potential, \textit{Forum
Math.} {\bf 8} (1996), 63--69

\bibitem{KW} Z. Katznelson, B.Weiss, A simple proof of some ergodic theorems, \textit{Israel J. Math.} {\bf 34} (1982), 291--296

%\bibitem{Kin} J.F. C. Kingman, The ergodic theory of subadditive stochastic p%rocesses, \textit{J.Royal Stat. Soc.}, {\bf B30}, (1968), 499--510

\bibitem{Kot} S.\ Kotani, Jacobi matrices with random potentials taking finitely many values, \textit{Rev.\ Math.\ Phys.} {\bf 1} (1989), 129--133

\bibitem{LP} J. C.  Lagarias, P.  A. B.  Pleasants,  Repetitive Delone
  Sets and Quasicrystals,   to appear in:  \textit{Ergod. Th. \& Dynam. Sys. }

\bibitem{LS} Y. Last, B. Simon, Eigenfunctions, transfer matrices, and absolutely continuous spectrum for one-dimensional Schr\"odinger operators, \textit{Invent. Math.} {\bf 135} (1999), 329--367

%\bibitem{Len0} D. Lenz, Aperiodische Ordnung und gleichm\"assige spektrale Ei%genschaften von Quasikristallen, Dissertation, Frankfurt/Main, Logos, Berlin (%2000)

\bibitem{Len1} D. Lenz, Random operators and crossed products, \textit{Mathematical Physics, Analysis and Geometry} {\bf 2} (1999), 197--220

\bibitem{Len2} D. Lenz, Uniform ergodic theorems on subshifts over a finite alphabet, to appear in: \textit{ Ergod. Th. \& Dynam. Sys.}

\bibitem{Len3} D. Lenz, Hierarchical structures in Sturmian dynamical systems, preprint

%\bibitem{Lot} M. Lothaire,  {\it Combinatorics on words}, Encyclopedia of
%  Mathematics and Its Applications, {\bf 17}, Addison-Wesley, Reading,
%  Massachusetts (1983)

%\bibitem{PSW} T. Poerschke, G. Stolz, J. Weidmann,  Expansions in generalized eigenfunctio%ns of selfadjoint operators, \textit{Math. Z.}, {\bf 202} (1989), 397--408


\bibitem{Que} M.  Queff\'{e}lec,   \textit{Substitution Dynamical
    Systems - Spectral Analysis}, Lecture Notes in Mathematics,
  Vol. 1284, Springer, Berlin, Heidelberg, New York (1987)

\bibitem{Sen} M. Senechal, \textit{Quasicrystals and geometry}, Cambridge University Press, Cambridge, (1995)

\bibitem{Sut} A.\ S\"ut\H{o}, The spectrum of a quasiperiodic Schr\"odinger operator, \textit{Commun.\ Math.\ Phys.} {\bf 111} (1987), 409--415

\bibitem{Sut2} A.\ S\"ut\H{o}, Singular continuous spectrum on a Cantor set of zero Lebesgue measure, \textit{J. Stat. \ Phys.} {\bf 56} (1989), 525--531

  
%\bibitem{Tes} G. Teschl, \textit{Jacobi operators and completely integrable nonlinear latt%ices}, Mathematical Surveys and Monographs, {\bf 72}, Amer. Math. Soc., Providence, R.I. (2%000)

\bibitem{Wal} P. Walters,  \textit{An introduction to ergodic theory}, Graduate Texts in Mathematics, {\bf 79}, Springer, Berlin (1982)


\end{thebibliography}
\end{document}